\documentstyle[preprint,aps]{revtex}
\begin{document}

\baselineskip 19pt

\title {The Extragalactic Neutrino Background Radiations From
        Blazars and Cosmic Rays}
\author{Arnon Dar
 and Nir J. Shaviv}
\address{Department of Physics and Space Research Institute,
Technion - Israel Institute of Technology, Haifa 32000, Israel}
\maketitle
\begin{abstract}

Blazar emission of gamma rays and cosmic ray production of gamma rays
in gas-rich clusters have been proposed recently
as alternative sources of
the high energy extragalactic diffuse gamma ray background radiation.
We show that these sources also produce a
very different high energy extragalactic
diffuse neutrino background radiation. This neutrino background
may be detected by the new generation of large neutrino telescopes
under construction and may be used to trace the origin of
the extragalactic gamma radiation.

\end{abstract}



In addition to the galactic diffuse gamma radiation, which varies
strongly with direction and can be explained by cosmic ray interactions
in the galactic interstellar medium\cite{BE}, there appears to be a
diffuse extragalactic gamma radiation which is isotropic at least
on a coarse scale \cite{EC}. Its existence has been confirmed recently
by analyses of observations with the  Energetic Gamma Ray Experiment
Telescope (EGRET) on board the Compton Gamma Ray Observatory
(CGRO) \cite{OE,HE,DE}. Various unresolved
extragalactic discrete and diffuse sources of gamma rays had been
suggested in the past to explain its origin, but all of them
have been later questioned by observations \cite{RM}.
Recently, however, attention has focused on two alternative sources
of the extragalactic diffuse gamma ray background radiation (GBR):
The detection of some active galactic nuclei (AGN) by EGRET
in high energy gamma rays, all belonging to the blazar type \cite{VM},
has led to the suggestion that blazar emission has produced
the extragalactic diffuse GBR \cite{CD}. On the other hand,
the discovery of very large mass of gas in intergalactic space within
groups and clusters of galaxies with the ROSAT and EINOBS X-ray
telescopes \cite{UGB} has led to the alternative suggestion
that cosmic ray interactions in groups and clusters
produced the extragalactic diffuse GBR \cite{DS}. In this paper
we show that these two alternative sources should have also produced
a diffuse extragalactic high energy neutrino background
radiation (NBR). These backgrounds are very different for the two
sources. They
may be detected by the new generation of large neutrino telescopes
under construction and may help to trace the origin of
the extragalactic diffuse GBR.

When it was first suggested that the diffuse extragalactic GBR
is the sum of gamma ray emission from unresolved AGN \cite{AW}
only the relatively nearby
quasar 3C 273 had been seen in high energy gamma rays\cite{BN}
$(E>100~MeV)$. Since the launch  of CGRO, 33 AGN were detected with
EGRET in high energy gamma rays\cite{VM},
all of which seem to belong to the blazar class. Many more
AGN including blazars, which are both bright and relatively close
and which were within the EGRET field of view, have not been detected
in high energy gamma rays, indicating that not all of these objects
are so luminous in gamma rays, or their emission is highly beamed or
has a low duty cycle. The beaming hypothesis is further supported by
other features of the sources such as superluminal velocity, beamed radio
emission, and a power-law gamma ray spectrum (with a power index between
1.4  and 3.0, with values between 1.8 and 2 being most common).
Although it is
difficult to see how the observed isotropic GBR can be produced
by the highly beamed and time variable emission from blazars, many of
which have a much  harder spectrum than that of the GBR, some authors
have actually shown that blazars with a plausible evolution function
could have produced the observed extragalactic GBR \cite{CD}.
If blazars produce their high energy beamed gamma rays via inverse
Compton scattering or via electron-positron annihilation in flight then
blazars are not significant cosmic sources of high energy neutrinos.
However, if blazars accelerate high energy cosmic ray nuclei that
collide with matter or radiation  and produce neutral pions
which decay into the observed high energy gamma rays, then
they also produce charged pions and kaons which decay into high energy
neutrinos \cite{BG}. Since the blazar luminosities in high energy gamma
rays are much larger than their luminosities in lower energy photons,
the gamma ray blazars cannot be ``hidden sources'' of high energy
neutrinos \cite{VB}. The summed emission of such blazars produces
both a diffuse extragalactic GBR and a diffuse extragalactic
NBR which are closely related.

The second  explanation of the extragalactic diffuse GBR was
motivated by the recent discoveries of large quantities of gas in
intergalactic space within groups and clusters of galaxies \cite{UGB}.
Most of the high energy gamma ray emission of our Milky Way (MW) galaxy
can be explained by cosmic ray interactions with its interstellar
gas \cite{BE,RM}.
But simple considerations show
that the summed contributions of gamma ray emission
from external galaxies falls short by
more than an order of magnitude\cite{SE,RM} in explaining
the observed flux \cite{EC,OE,HE,DE} of the GBR \cite {PC}.
However, whereas the ratio of the total mass of gas to light in the MW is
only \cite{VDB}  $M_{gas}/L_{MW}\approx 4.8\times 10^9M_\odot/
(2.3\pm 0.6)\times 10^{10}L_\odot \approx 0.21~M_\odot/L_\odot$,
recent X-ray observations of groups and clusters of galaxies
have shown that they contain much larger mass of intergalatic gas
than their total stellar mass \cite{UGB}.  For instance,
analyses of recent observations
with the ROSAT X-ray telescope of the compact group HCG62 and the Coma
cluster yielded, $M_{gas}/L_B\approx 4.4\times 10^{11}M_\odot
h^{-5/2}/ 2.4\times 10^{10}h^{-2}L_\odot
\approx 19h^{-1/2} M_\odot/L_\odot$, within a distance of
$0.24h^{-1}~Mpc$ from the center of HCG62\cite{TJ}, and
$M_{gas}/L_B\approx (5.45\pm 0.98) \times 10^{13}M_\odot
h^{-5/2}/1.8\times 10^{12}h^{-2}L_\odot
\approx (30\pm 6) h^{-1/2} M_\odot/L_\odot$,
within a distance of $1.5h^{-1}~Mpc$ from the center
of the Coma cluster\cite{UG}. It was also found  that these
ratios are rather typical to the groups and clusters of galaxies that
have been detected in X-rays and for large enough radii are
independent of radius. It was further argued on theoretical
grounds
that these ratios are universal in groups and clusters \cite{UGB}
and they imply that most of the baryonic matter in the Universe, as
estimated from Big-Bang Nucleosynthesis, is in intergalactic gas within
groups and clusters of galaxies \cite{DS}. This has been used by Dar and
Shaviv \cite{DS} to show that an average cosmic ray flux in
intergalactic space within clusters and groups similar to the
{\it average} cosmic ray flux observed in the MW \cite{SMC},
could have produced the extragalactic GRB by interactions
with the intergalactic gas in groups and clusters \cite{HW}.
Such a universal cosmic ray flux in groups
and clusters should have also produced high-energy diffuse galactic
and extragalactic neutrino background radiations which are closely
related to the diffuse galactic and extragalactic GBR.

The main mechanism by which cosmic rays produce high energy gamma rays
is by $\pi^0$ production in inelastic collisions with gas nuclei and/or
background photons followed by immediate $\pi^0 \rightarrow
2\gamma$ decay.
For simplicity we will assume that the target particles are gas nuclei
(the results are very similar if the production takes place on
background photons). The same interactions produce charged pions, kaons
and small quantities of other mesons which decay into electrons
and neutrinos
(e.g., $\pi \rightarrow \mu \nu_\mu$, $K \rightarrow \mu \nu_\mu$,
$\mu \rightarrow e \nu_e \nu_\mu$) if their  mean
free path for interaction is much larger than their decay path,
$\lambda_d=\gamma c \tau$, with $\gamma\equiv E/m_0 c^2$.
Because of the low mean baryonic density of the galactic and
intergalactic gas,
$ \lambda_d n_b \sigma_{in}\ll 1$, and all the secondary
unstable particles
produced in cosmic ray collisions decay before interacting with other
gas nuclei. For a power law spectrum of cosmic ray nuclei,
$dF_{CR}/dE\approx AE^{-p}$, and inclusive
cross sections for meson production that obey Feynman scaling,
the produced fluxes of $\gamma$-rays, $\nu_e$'s, $\nu_\mu$'s and
$\nu_\tau$'s are all proportional at high energies to the flux
of cosmic ray nuclei,
\begin{equation}
{dF_{\gamma}\over dE}\propto
{dF_{\nu_i}\over dE} \propto {dF_{CR} \over dE} ~.
\end{equation}
In particular, if $\nu_\mu + \bar{\nu}_\mu$
production proceeds mainly via $\pi$, $K$ and $\mu$ decays then  one
can use the analytical methods developed, e.g., by Dar \cite{AD} and by
Lipari \cite{LP} to show that the neutrino and gamma
ray fluxes produced by cosmic rays are related through
\begin{equation}
{dF_{\nu_\mu} \over dE}\approx 0.70 {dF_\gamma \over dE}~.
\end{equation}
Relation (2) can be verified also by detailed Monte Carlo calculations.
It is valid as long as the $\gamma
$'s are not absorbed in the intergalactic space
through the process $\gamma +\gamma \rightarrow e^+ + e^-$.

We can now use Eq. (2) to estimate the extragalactic diffuse NBR
at very high energies from the observed extragalactic diffuse GBR
in the GeV region \cite{EC,OE,HE,DE},
$dF_\gamma/ dE\approx 2\times
10^{-6}E^{-2.1\pm 0.1} ~
\left[cm^{-2}s^{-1}ster^{-1}GeV^{-1}\right]~:
$

The power law gamma ray
spectra of the 33 blazars which were detected by EGRET, if extrapolated
to TeV energies, suggest that many of them should have been easily
detected with currently available TeV gamma ray telescopes \cite{TC}.
To date only Markarian 421, the nearest (at redshift z=0.03)
of the AGN seen by EGRET, was observed in TeV gamma rays
with the Whipple Observatory gamma ray telescope
\cite{LE}. This has been interpreted as being due to the absorption
of TeV gamma rays from distant blazars
by the extragalactic IR background radiation \cite{FS}.
The spectral index of Markarian 421, which is
implied by the combined EGRET and Whipple observations
is $p=2.06\pm 0.05$.
If this is the typical power index of the high energy
gamma ray emission by blazars then the extragalactic diffuse
GBR must have this power index below 500 GeV where gamma ray
absorption in intergalactic space is negligible. This power index
is consistent with the best fitted \cite{OE,HE,DE}
power index of the extragalactic diffuse GBR which has been observed
so far at energies below 10 GeV.
If the TeV gamma rays
are produced by pion decay then the observed flux level of the
extragalactic diffuse GBR implies the existence of an extragalactic
diffuse NBR with a flux
\begin{equation}
{dF_ {\nu_\mu}\over dE} \sim 1.4\times
10^{-6}E^{-2.06\pm0.05} ~
\left[cm^{-2}s^{-1}ster^{-1}GeV^{-1}\right],
\end{equation}
which extends at least up to energies of a few TeV.
Without the knowledge of the gamma ray emission of blazars beyond
10 TeV it is not possible to predict reliably the extragalactic
GBR and NBR beyond 10 TeV which are produced by blazars.

If the extragalactic diffuse GBR is
produced by a universal cosmic ray flux in groups and clusters,
with an average flux similar to that observed in the Milky Way,
then the power index of the GBR above 10 GeV must change to  2.7,
which is the power index of high energy cosmic rays below $10^4$
TeV. Such a cosmic ray flux must have also produced an extragalactic
diffuse NBR which extends all the way to the highest cosmic ray
energies. Its flux below $10^4$ TeV is given by
\begin{equation}
{dF_{\nu_\mu} \over dE} \approx 1.4\times 10^{-6}E^{-2.7}
\left[cm^{-2}s^{-1}ster^{-1}GeV^{-1}\right].
\end{equation}

Eq. (2) and the observed diffuse galactic GBR
can  be used also to evaluate the diffuse galactic NBR.
Whereas the diffuse galactic gamma radiation depends on galactic
coordinates,
reflecting variations in the local intensity of cosmic rays and  in the
density of interstellar gas in the MW, the extragalactic diffuse
gamma radiation seems to be isotropic.

At low energies both
the galactic and extragalactic diffuse NBR are masked by the
atmospheric neutrino background.
At high energies the
atmospheric neutrino flux is dominated by $\pi$ and
$K$ decays since the contibution from $\mu$ decays is strongly suppressed
by an additional power of E. For $E>> 1 TeV$ and zenith angels not too
close to the horizon, it is given approximately by \cite{AD,LP}
\begin{equation}
{dF_{\nu_\mu} \over dE} \approx 5.3~sec~\theta~E^{-3.7}~
\left[cm^{-2}s^{-1}ster^{-1}GeV^{-1}\right].
\end{equation}
In Fig. 1  we plotted our predictions for
the atmospheric NBR at zenith angles $\theta=0^0,~ 90^0$,
 the galactic NBR and the extragalactic
NBR produced by blazars and by cosmic rays, respectively. As can be seen
from Fig. 1, an extragalactic NBR produced by blazars dominates
the atmospheric NBR already at 10 TeV while an extragalactic
NBR produced by a universal cosmic ray flux in groups and clusters
dominates the atmospheric NBR only at energies above
$\sim 4\times 10^3 TeV$.

In principle the extragalactic NBR can be distinguished
from the galactic and atmospheric NBR
because it is isotropic, while the atmospheric neutrino flux at
high energies depends on zenith angle
and the galactic neutrino flux depends on galactic coordinates:
The galactic NBR is non isotropic even at very high energies because
it is proportional to the column density of gas in the MW as seen from
the solar system ($\sim 8.5 kpc$ away
from the center of the MW galaxy) in different
directions. The atmospheric neutrino
background depends on zenith angle because the probability of
very energetic pions and kaons to decay and produce neutrinos
before being absorbed in the atmosphere depends on zenith angle
(see Eq. (5) and Refs. \cite{AD,LP}).
The predicted blazar produced extragalactic NBR
is detectable by the large neutrino telescopes under construction
\cite{FH}. The predicted
extragalactic NBR produced by cosmic ray interactions
in groups and clusters is detectable only by the future generations
of large ($>1~km^2$) neutrino telescopes \cite {JL}.

It is quite possible that we have underestimated the atmospheric
neutrino background at very high energies because we have neglected
the contribution from production and decay of heavy flavour mesons. At
energies below 1 TeV, the inclusive cross sections for production of
charm, beauty and truth in proton-proton and proton-nucleus collisions
are much smaller than for normal and strange meson production.
Therefore $\pi$, and $K$ production and decay dominate there the
atmospheric NBR. However, at very high energies $\pi$
and $K$ decays in the atmosphere are strongly suppressed  and the
production of charmed and beauty mesons which decay promptly
may dominate the atmospheric
meutrino flux.  Unfortunately, the present experimental
information on heavy flavour production and decay cannot
be extrapolated reliably to very high energies where their
contribution may dominate the atmospheric NBR.

Many other exotic sources that may have generated an extragalactic
high energy diffuse NBR have been proposed by various authors
\cite{VB}. In principle, the extragalactic NBR produced by them
can be distinguished from the NBR produced by the above conventional
sources, by their flux levels, their
spectra, and their typical spatial and temporal variabilities.
\bigskip
{\bf Acknowledgement}: We thank J.N. Bahcall for proposing to us
to calculate the
extragalactic NBR produced by cosmic ray interactions
in intergalactic space within clusters and groups of galaxies.
\newpage
\centerline{\bf References and Footenotes}

\centerline{\bf Figure Caption}
\begin{figure}
\noindent{ Fig. 1. Comparison between the predicted
NBR produced by blazars (dotted line),
by a universal MW-like cosmic ray flux
in groups and clusters (dashed line), by cosmic rays in the MW galaxy
from the direction of the galactic center (dashed-doted line) and the
predicted atmospheric neutrino background at $0^0$ and $90^0$ zenith
angles (full lines).}
\end{figure}
\end{document}